\documentclass[doublecol]{epl2} 
\usepackage[applemac]{inputenc}
\usepackage[T1]{fontenc} 
 \usepackage{amsmath} 
\usepackage{subfigure}
\usepackage{lipsum}
\usepackage{flushend}
\usepackage{amsfonts} 
\usepackage{amssymb, mathrsfs}
\usepackage{braket}
\usepackage{graphicx} 

\usepackage{bbm}

\def\beq{\begin{equation}}
\def\eeq{\end{equation}}
\def\bsp{\begin{split}}
\def\esp{\end{split}}
\def\bea{\begin{eqnarray}}
\def\eea{\end{eqnarray}}
\def\ba{\begin{array}}
\def\ea{\end{array}}

\def\lb{\left(}
\def\rb{\right)}

\def\l.{\left.}
\def\r.{\right.}

\def\ra{\rangle}
\def\la{\langle}

\title{Magnonic triply-degenerate nodal points}
\shorttitle{Magnonic triply-degenerate nodal points} 

\author{S. A. OWERRE\inst{1} }

\shortauthor{S. A. OWERRE}

\institute{                   
  \inst{1} Perimeter Institute for Theoretical Physics - 31 Caroline St. N., Waterloo, Ontario N2L 2Y5, Canada\\
}
\pacs{75.30.Ds}{Spin waves}

\abstract{
We generalize the concept of  triply-degenerate nodal points to  non-collinear  antiferromagnets.       Here, we introduce this concept to insulating quantum antiferromagnets on the decorated honeycomb lattice, with spin-$1$ bosonic quasiparticle excitations known as magnons.  We demonstrate the existence of magnonic surface states with constant energy contours that form pairs of magnonic arcs connecting the surface projection of the magnonic triple nodal points.  The quasiparticle excitations near the triple nodal points represent three-component bosons beyond  that of   magnonic Dirac, Weyl,  and nodal-line cases. They can be regarded as a direct reflection of the intrinsic spin carried by magnons. Furthermore, we show that  the magnonic triple nodal points can  split into magnonic Weyl points,  as  the system  transits from a non-collinear spin structure  to a noncoplanar one with a nonzero scalar spin chirality.  Our results not only apply to insulating antiferromagnets, but also provide a platform to seek for triple nodal points in metallic antiferromagnets.
}

\begin{document}

\maketitle

\textbf{Introduction.}-- In high-energy physics, relativistic  symmetry imposes a constraint  on the varieties of fermions allowed.  The three known relativistic fermions in high-energy physics are  Dirac, Weyl, and Majorana fermions.  The recent excitement in condensed-matter physics is  that the fermionic elementary particles can be realized as low-energy  topologically protected gapless electronic  quasiparticle excitations in solid-state materials.  They are   termed  topological semimetals, and possess unusual properties such as open Fermi arc surface states. In recent years, topological Dirac and Weyl semimetals have been theoretically proposed \cite{dsm1,dsm2,wsm0,wsm1,wsm2} and experimentally confirmed \cite{dsm3,dsm4,dsm5,wsm3,wsm4}.  

In the topologically protected Dirac semimetals, crystal and time-reversal symmetry guarantee that a pair of doubly-degenerate electronic bands cross linearly at isolated points in momentum space; whereas  Weyl semimetals are realized in systems with either broken inversion or time-reversal symmetry, with only two non-degenerate linear-band crossing at isolated points in momentum space. Around the band crossing points near the Fermi energy, the low-energy quasiparticle excitations behave as Dirac or Weyl  fermions, and are described by Dirac or Weyl equation.  The topological protection of the linear band crossings have distinctive Fermi arcs which connect projected bulk Dirac or Weyl points on the surface Brillouin zone (BZ).  In addition, solid state materials can also allow  two linear-band crossing  along a one-dimensional line or ring as opposed to isolated points, and they are termed nodal-line semimetals \cite{nsm,nsm1}.

As condensed-matter quasiparticle excitations are not constrained by relativistic  symmetry, there is a possibility to realize other types of fermionic quasiparticle excitations with no high-energy physics analogs. Recently, both theory and experiment have realized symmetry-protected condensed-matter quasiparticle excitations with 3-, 6-, and 8-fold degenerate band crossings \cite{tp0,tp1,tp2,tp3,tp4,tp5,tp6,tp6a, tp6b, tp7,tp8, tp8a}. Of particular interest is the 3-fold band degeneracy known as the triply-degenerate nodal points (TPs). They are formed by the crossing of one doubly-degenerate band and one single non-degenerate band along the high symmetry lines in momentum space. They are also described by three-component fermions, and therefore differ from the band crossings in Dirac, Weyl, and nodal-line semimetals.  

Indeed, the topological aspects of band structures in condensed-matter systems  are independent of the statistical nature of the quasiparticle excitations.  Recently, researchers have come to the realization that topologically protected linear band crossings in condensed-matter systems are not limited to fermionic quasiparticle excitations, but can also occur in bosonic quasiparticle excitations. In this case the quasiparticle excitations obey the Bose-Einstein stastistics, therefore linear band crossing must occur at finite energy with the lowest band being the most realistic due to the population effect of bosons at low temperatures. In fact, bosonic Weyl points have been experimentally observed in photonic \cite{lu} and phononic \cite{fee} crystals. These concepts have also been generalized to insulating magnetic systems, which have spin-$1$ bosonic quasiparticle excitations called magnons. Thus far, gapless topologically protected magnonic Dirac \cite{md1,md2,md3,md4}, Weyl \cite{mw1, mw2, mw3, mw4, mw5, mw6, mw7, mw8}, and nodal-line \cite{nw0,nw1} semimetals have been theoretically proposed in insulating quantum magnets.  

In this Letter, we generalize the concept of  TPs to insulating  non-collinear  quantum antiferromagnets on the decorated honeycomb lattice, which is also known as the star lattice \cite{zheng, rich}. We show the existence of magnonic TPs in the  120$^\circ$ non-collinear spin structure, protected by magnetic crystal symmetry. The bosonic quasiparticle excitations near the magnonic TPs represent three-component bosons beyond the two-band crossing points in previously studied magnonic  Dirac, Weyl, and nodal-line cases. We also show the existence of magnonic  arcs connecting the surface projection of the magnonic TPs.  Interestingly, most of  the currently known  electronic TPs are non-magnetic \cite{tp0,tp1,tp2,tp3,tp4,tp5,tp6, tp6a, tp6b, tp7,tp8, tp8a}, hence our results not only apply to insulating antiferromagnets, but may also  pave the way to search for TPs in metallic  antiferromagnets with (non)collinear spin structures.

 \begin{figure*}
\centering
\includegraphics[width=.75\linewidth]{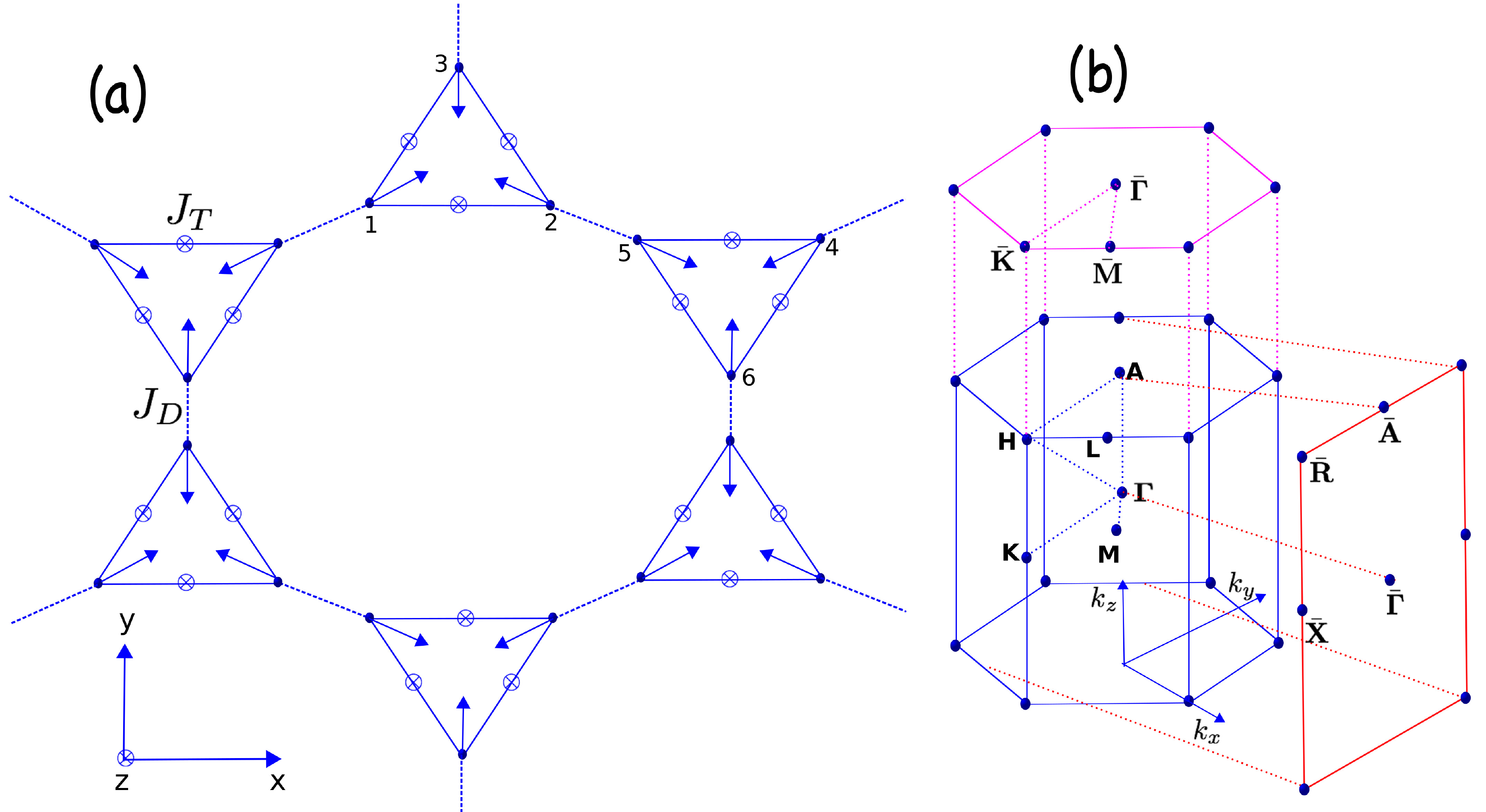}
\caption{Color online. (a) Top view of  unshifted decorated honeycomb lattice stacked along the (001) direction. The decorated honeycomb lattice has two topologically inequivalent antiferromagnetic nearest-neighbour (NN) bonds $J_D$ (dimer  bonds) and $J_T$ (triangular bonds).  The unit cell is indicated by numbers and the $120^\circ$ non-collinear spin configuration with a positive vector chirality is indicated  by the blue arrows. The direction of the DMI along the (001) direction (out-of-plane)  is denoted by the dotted circles. (b) The bulk Brillouin zone (BZ)  and  its projection onto the hexagonal (001) and rectangular (010) surface BZ. }
\label{lattice}
\end{figure*}

 \textbf{Spin model.}-- We consider the microscopic  spin Hamiltonian of non-collinear antiferromagnets on the frustrated decorated honeycomb lattice
\begin{align}
\mathcal H&=J_T\sum_{\la ij\ra,\ell} {\bf S}_{i,\ell}\cdot{\bf S}_{j,\ell} + J_D\sum_{\la ij\ra,\ell} {\bf S}_{i,\ell}\cdot{\bf S}_{j,\ell}\label{p1}\\&\nonumber+\sum_{\la ij\ra,\ell}\bold{D}_{ij}\cdot {\bf S}_{i,\ell}\times{\bf S}_{j,\ell}+ J_\perp\sum_{i,\la \ell \ell^\prime\ra}{\bf S}_{i,\ell}\cdot{\bf S}_{i,\ell^\prime},
\end{align}
where $i$ and $j$ denote the sites on  the  layers, $\ell$ and $\ell^\prime$ label the layers.  The first two terms are  topologically inequivalent antiferromagnetic nearest-neighbour (NN) bonds $J_D$ (dimer bonds) and $J_T$ (triangular  bonds) on the decorated honeycomb lattice (see Fig.~\ref{lattice}).  The third term is an out-of-plane antisymmetric  Dzyaloshinskii-Moriya interaction  (DMI) \cite{dm1,dm2}  $\bold D_{ij}= - D_z \hat{z}$. The DMI is a consequence of spin-orbit coupling (SOC), and it is present in magnetic systems that lack  inversion symmetry between two magnetic spin sites on each layer. Without loss of generality, we  consider $120^\circ$ non-collinear spin structure with positive vector chirality, which is stabilized by $D_z>0$.  The last term is the interlayer coupling, which can be ferromagnetic ($J_\perp<0$) or antiferromagnetic ($J_\perp>0$). In this paper, we consider antiferromagnetic interlayer coupling. The ferromagnetic one can be trivially obtained.

   In principle, all the interactions in Eq.~\eqref{p1} are intrinsic to realistic magnetic materials with a decorated honeycomb lattice structure \cite{zheng}.  Due to the geometry of the decorated honeycomb lattice, the limit $J_D>J_T$ is isomorphic to a honeycomb lattice where the dimer bonds dominate, whereas the opposite limit $J_T>J_D$ is isomorphic to a kagome lattice where the triangular bonds dominate. Therefore, the decorated honeycomb lattice  can be visualized as an interpolating lattice between the honeycomb and kagome lattices. Moreover, it can also be considered as the parent system  from which magnonic topological semimetals can be realized in the honeycomb and kagome antiferromagnets.  Nevertheless, the decorated honeycomb lattice has a larger unit cell (with six sites)  than both the honeycomb and kagome lattices. Therefore it possesses a richer band structure. Also note that  magnonic TPs cannot occur in the conventional honeycomb lattice as there are only two magnon bands.

 \textbf{Magnonic triply-degenerate nodal points.}--
  We first present a synopsis of the symmetry analysis of the system.   We consider  $120^\circ$ non-collinear spin structure on the decorated honeycomb lattice stacked congruently  along the (001) or $z$ direction. In other words, there is a non-negligible interlayer coupling which is always present in realistic magnetic materials \cite{zheng}, and leads to quasi-two-dimensional (quasi-2D) or three-dimensional (3D) spin structures. Due to intrinsic  DMI, the inversion symmetry of the lattice is broken. Nevertheless, the system possesses $\mathcal C_{3}$, $\mathcal C_{6}$, $\mathcal M_y$, and $\mathcal M_z$ symmetries. Here, $\mathcal C_{3}$ is three-fold rotation symmetry along the $z$ direction,  $\mathcal C_{6}$ is $\pi/3$ rotation with respect to the center of a dodecagon, $\mathcal M_y$ denotes the mirror reflection symmetry about the $x$-axis that sends $y\to-y$, and $\mathcal M_z$ denotes the mirror reflection symmetry about the $x$-$y$ plane that sends $z\to-z$. All symmetries preserve the lattice and the magnetic order expect for  $\mathcal M_y$ which reverses the magnetic order. Therefore, under time-reversal symmetry ($\mathcal T$) the combined operation $\mathcal T\mathcal M_y$ is also a symmetry of the system. 

\begin{figure}
\centering
\includegraphics[width=1\linewidth]{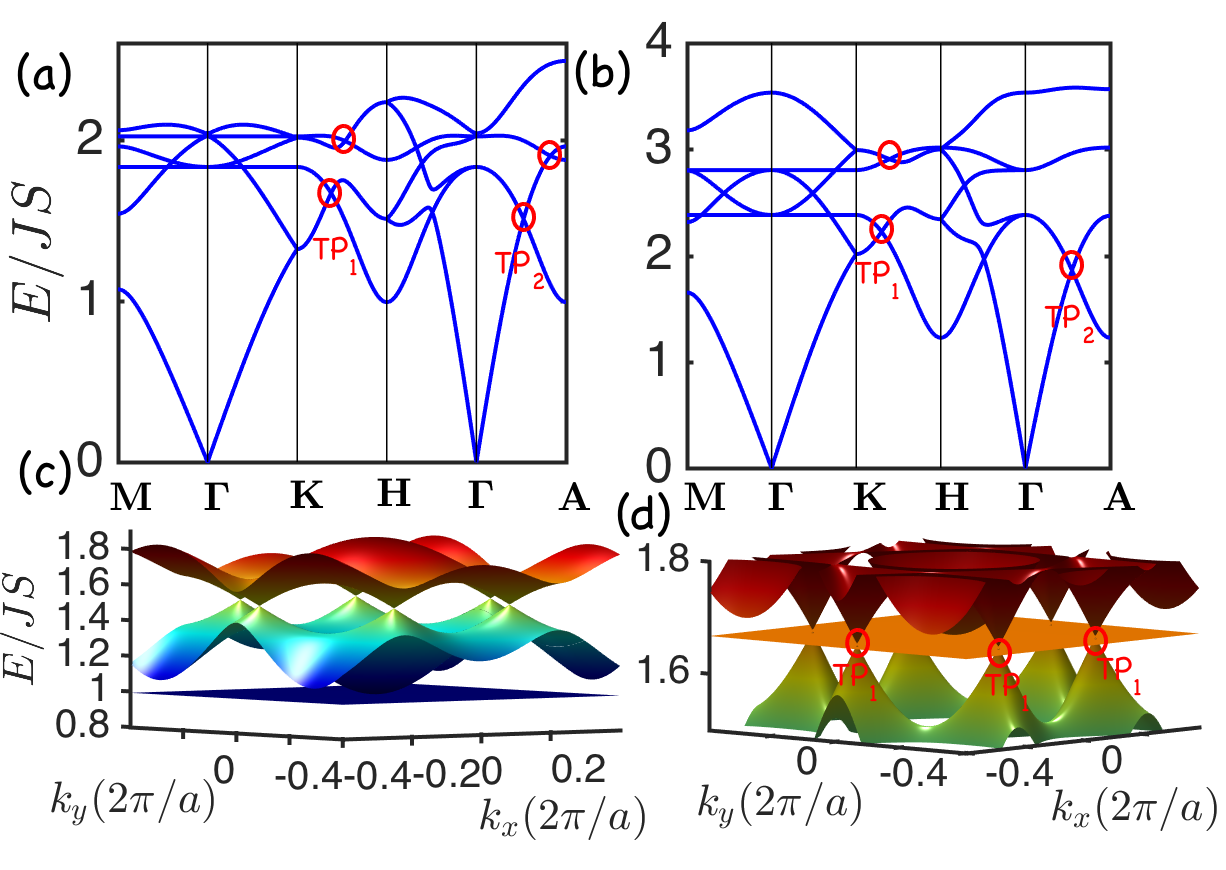}
\caption{Color online.  Magnon band dispersion  of $120^\circ$ non-collinear spin structure (with zero scalar spin chirality) on the stacked decorated honeycomb lattice. (a)    $J_D/J_T = 0.5$,  $D_z/J_T = 0.15$, $J_\perp/J_T = 0.45$.  (b)   $J_D/J_T = 1.5$,  $D_z/J_T = 0.15$,  $J_\perp/J_T = 0.45$. The red circles highlight the TPs formed by the crossing of one doubly-degenerate band and one single non-degenerate band along  ${\bf K}$-${\bf H}$  and  ${\bf \Gamma}$-${\bf A}$ lines. (c) and (d) depict the 3D magnon band dispersions in the $k_x$-$k_y$ plane for  constant $k_z$ values  above $(k_z=0.5\times 2\pi/c)$ and at $(k_z=k_z^{\text{TP}_1}=0.1898\times 2\pi/c)$ the magnonic triple point (TP$_1$) respectively.   The flat band corresponds to a lifted zero-energy mode due to the DMI,  and can  acquire a small dispersion by the inclusion of a next-nearest neighbour  antiferromagnetic interaction.}
\label{TDNLM}
\end{figure}

Now, we note that the decorated honeycomb lattice has six sites in the unit cell (see Fig.~\ref{lattice}(a)), therefore we expect six magnon bands (see Supplemental Material) along each high symmetry line of the Brillouin zone (BZ) (see Fig.~\ref{lattice}(b)).  As shown in Figs.~\ref{TDNLM}(a) and (b), there are different magnon band crossing points along the high symmetry lines in momentum space.   The type of magnon band crossing points is related to the  symmetry protection  of the system. For instance,  there are several magnonic nodal-lines (NLs) and Dirac points (DPs) (i.e. linear crossing of two non-degenerate bands) on the $k_z=0$  mirror plane (${\bf M}$-${\bf \Gamma}$-${\bf K}$ lines), as well as $k_z=\pi$ plane (not shown) and ${\bf H}$-${\bf \Gamma}$ line.  The non-degenerate linear band crossing points are protected by  $\mathcal T\mathcal M_y$ symmetry and can be described by a two-component Hamiltonian in the vicinity of the linear band crossing. 

   Quite distinctively, there are different magnon band crossing points  along ${\bf K}$-${\bf H}$  and  ${\bf \Gamma}$-${\bf A}$ high-symmetry lines. The new linear band crossing points are different from  those along ${\bf M}$-${\bf \Gamma}$-${\bf K}$ and ${\bf H}$-${\bf \Gamma}$ lines in that they involve three magnon bands that cross  each other simultaneously at a common nodal point. This leads to a magnonic TP degeneracy. Note that they  are formed by the crossing of one doubly-degenerate band and one single non-degenerate band along  ${\bf K}$-${\bf H}$  and  ${\bf \Gamma}$-${\bf A}$ high-symmetry lines. Next, we consider the perpendicular planes intersecting the magnonic TPs (i.e. the $k_x$-$k_y$ planes for constant $k_z$ values). For the planes above the magnonic TPs,  two of the three non-degenerate magnon bands  touch at  isolated  points as shown in  Fig.~\ref{TDNLM}(c). Whereas for the planes exactly at the magnonic TPs, the three non-degenerate magnon bands touch simultaneously and linearly at the same point as shown in Fig.~\ref{TDNLM}(d) (see also Supplemental Material).  The flat magnon band in Figs.~\ref{TDNLM}(c) and (d) corresponds to a lifted zero-energy mode due to the DMI \cite{dm3}.  It can  acquire a small dispersion by the inclusion of a next-nearest neighbour antiferromagnetic interaction (not shown), but this interaction does not remove the magnonic TPs as it breaks no symmetry. The magnonic TPs are protected by $\mathcal C_{3}$ and $\mathcal M_{z}$ symmetry.  It should be noted that the existence of protected magnonic TPs strictly requires the presence of  quasi-2D or 3D spin structures, i.e. nonvanishing interlayer coupling.  The effective Hamiltonian near the magnonic TPs has the form \cite{tp1,foot}.
\begin{align}
\mathcal H_{\text{TP}}= E_{\text{TP}}\bold{I}_{3\times 3}+\sum_{i=x,y,z}v_iq_i\lambda_i,
\end{align}
where $\bold{q}=\bold{k}-\bold{k}_{\text{TP}}$ is the momentum deviation from the crossing point at $\bold{k}_{\text{TP}}$, $E_{\text{TP}}$ is the energy of the TPs, $v_i$ are the group velocities, and $\lambda_i$  are the $3\times 3$ spin-$1$  matrix representations, whose explicit forms are model dependent. The magnonic TPs are a direct reflection of the intrinsic spin carried by magnons. A linear Hamiltonian of this form carries a Chern number of $\mathcal C=-2,0,2$, which is only well-defined when a gap opens \cite{foot1, foot2, foot3}.

 \textbf{Magnonic Weyl points.}-- Magnonic Weyl points (WPs) can be generated from TPs by breaking of symmetry.  There are two ways to macroscopically break symmetry in this model. The first one is to allow an in-plane DMI, which  breaks U(1) rotation invariance and induces spin canting out-of-plane (noncoplanar spin structure) with a nonzero scalar spin chirality given by $\chi_{ijk;l}=  {\bf S}_{i,\ell}\cdot\lb {\bf S}_{j,\ell}\times{\bf S}_{k,\ell}\rb$. The scalar chirality macroscopically breaks $\mathcal T$ symmetry as well as the combined symmetry $\mathcal T\mathcal M_y$. Note that the noncoplanar spin structure also breaks $\mathcal M_{z}$ symmetry, but $\mathcal C_{3}$ symmetry is preserved.  The second one is to apply an external magnetic field along the (001) direction. This also leads to similar spin canting or noncoplanar spin arrangements  as the in-plane DMI, and therefore breaks the same symmetries. Here, we consider  noncoplanar spin structures induced by an external magnetic field along the (001) direction (see Supplemental Material). In this scenario, the magnonic NLs and DPs along ${\bf M}$-${\bf \Gamma}$-${\bf K}$ and ${\bf H}$-${\bf \Gamma}$ lines will be completely gapped out due to broken $\mathcal T\mathcal M_y$ symmetry in the noncoplanar regime.

\begin{figure}
\centering
\includegraphics[width=1\linewidth]{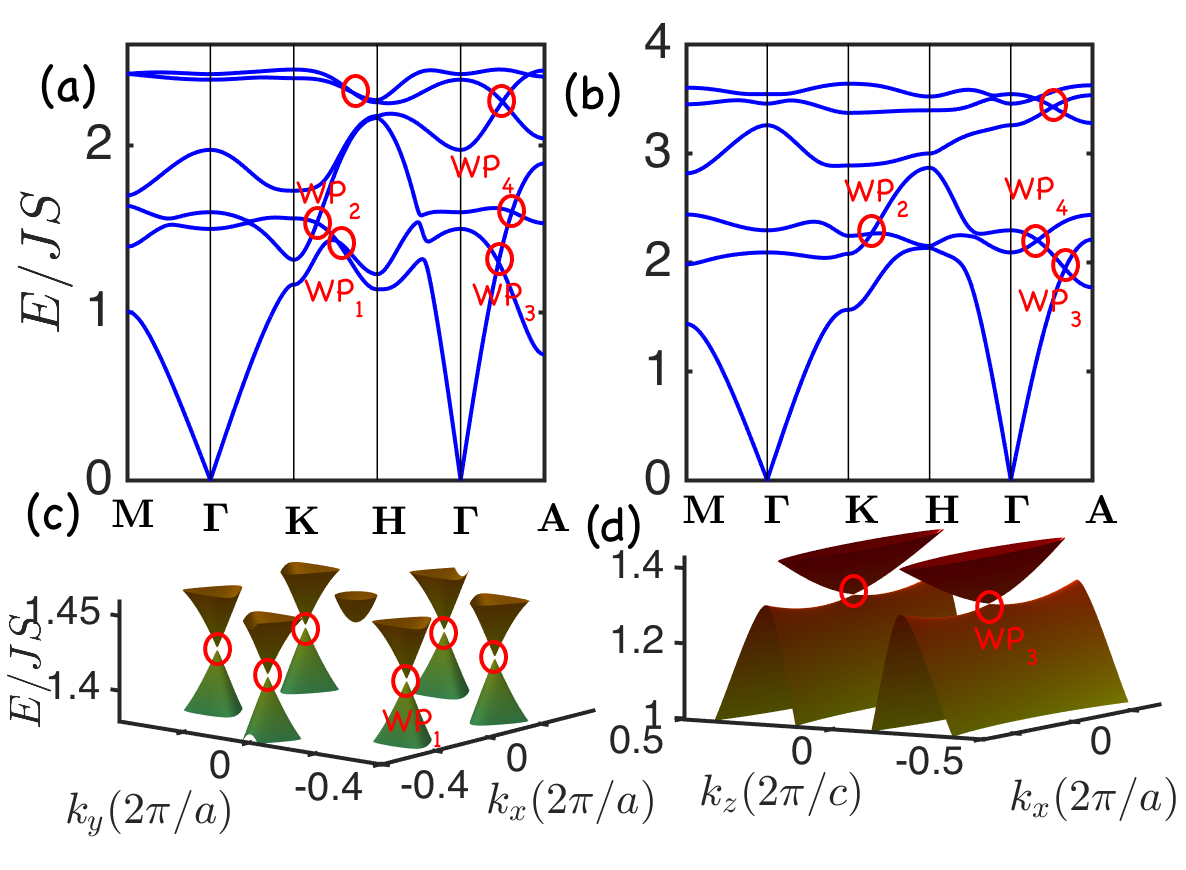}
\caption{Color online. Magnon band dispersion  of noncoplanar spin structure (with nonzero scalar spin chirality) on the stacked decorated honeycomb lattice.  The red circles highlight the magnonic WPs formed by the crossing of two non-degenerate bands along  ${\bf K}$-${\bf H}$  and  ${\bf \Gamma}$-${\bf A}$ lines. (a)    $J_D/J_T = 0.5$,  $D_z/J_T = 0.15$, $J_\perp/J_T = 0.45$, $H=0.3H_s$.  (b)   $J_D/J_T = 1.5$,  $D_z/J_T = 0.15$,  $J_\perp/J_T = 0.45$, $H=0.3H_s$. (c) and (d) depict the 3D Weyl cones  in the $k_x$-$k_y$ plane for WP$_1$ at  $k_z=0.262\times 2\pi/c$  and in the $k_x$-$k_z$  plane  for WP$_3$  at $k_y=0$  respectively.  }
\label{WM}
\end{figure}

In contrast, the preserving of $\mathcal C_{3}$ symmetry in the noncoplanar regime means that the magnonic TPs cannot be fully gapped out. Rather, they will split into magnonic WPs  along ${\bf K}$-${\bf H}$  and  ${\bf \Gamma}$-${\bf A}$ lines, where two non-degenerate magnon bands cross linearly at isolated points.    As shown in Fig.~\ref{WM}(a) and (b), the magnon bands are gapped out everywhere except along  ${\bf K}$-${\bf H}$  and  ${\bf \Gamma}$-${\bf A}$ lines, where two non-degenerate bands cross linearly at the WPs.  The lowest magnon band has a total of six magnonic WPs (four along $(\pm){\bf K}$-${\bf H}$  and  two ${\bf \Gamma}$-$(\pm){\bf A}$) in the entire BZ with opposite chirality. In fact, this is one of the most interesting property of this model --- that the magnonic WPs occur at the lowest excitation. The effective Hamiltonian near the WPs has the same 2-component form 
\begin{align}
\mathcal H_{\text{WP}}= E_{\text{WP}}\bold{I}_{2\times 2}+\sum_{i=x,y,z}v_iq_i\sigma_i,
\label{DP}
\end{align}
where $\bold{q}=\bold{k}-\bold{k}_{\text{WP}}$ is the momentum deviation from the crossing point at $\bold{k}_{\text{WP}}$, $E_{\text{WP}}$ is the energy of the WPs, $v_i$ are the group velocities and $\sigma_i$ are the $2\times 2$ Pauli matrices. But in this case the magnon bands carry a Chern number of $\mathcal C= \pm\text{sgn}(\sin\phi)=\pm 1$, where $\phi$ is the angle subtended by three noncoplanar spins in a unit triangle (see Supplemental Material), and $\sin\phi$ is proportional to  $\chi_{ijk; \ell}$.  Although  bosonic TPs and WPs must occur at finite energy, the bosonic WPs in the lowest excitation have been proven to be the most realistic as reported experimentally in photonic \cite{lu} and phononic  \cite{fee} crystals. Therefore the current magnonic TPs and WPs are within experimental reach in realistic magnetic materials \cite{zheng}. 
 
 \begin{figure}
\centering
\includegraphics[width=1\linewidth]{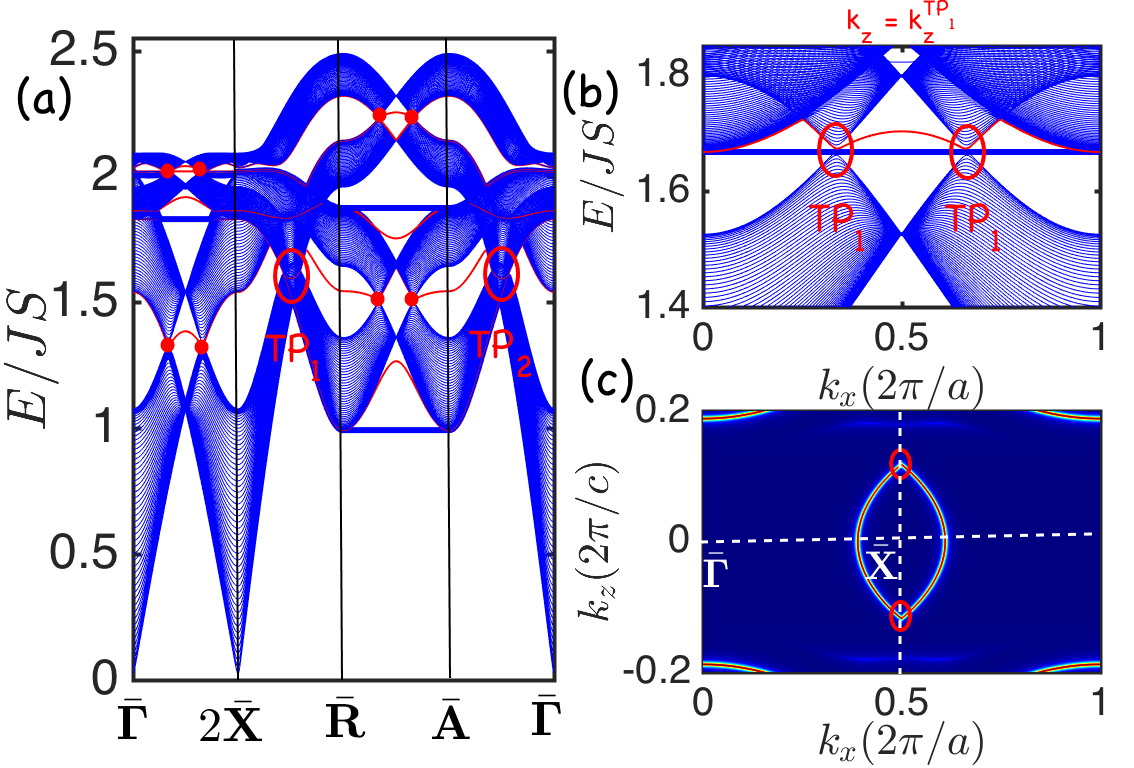}
\caption{Color online.  (a) The projected (010)-surface dispersion along the surface BZ. The red circles are the projected TPs, and filled red circles are the projected DPs. The  red curves indicate the surface states.    (b) The projected (010)-surface dispersion along $k_x$ direction with constant $k_z=k_z^{\text{TP}_1}=0.1898\times 2\pi/c$ at TP$_1$. (c) Magnonic arcs with energy set at TP$_1$  for the (010)-surface state.   The parameters are the same as  in Fig.~\ref{TDNLM}(a).}
\label{TSS}
\end{figure}

\begin{figure}
\centering
\includegraphics[width=1\linewidth]{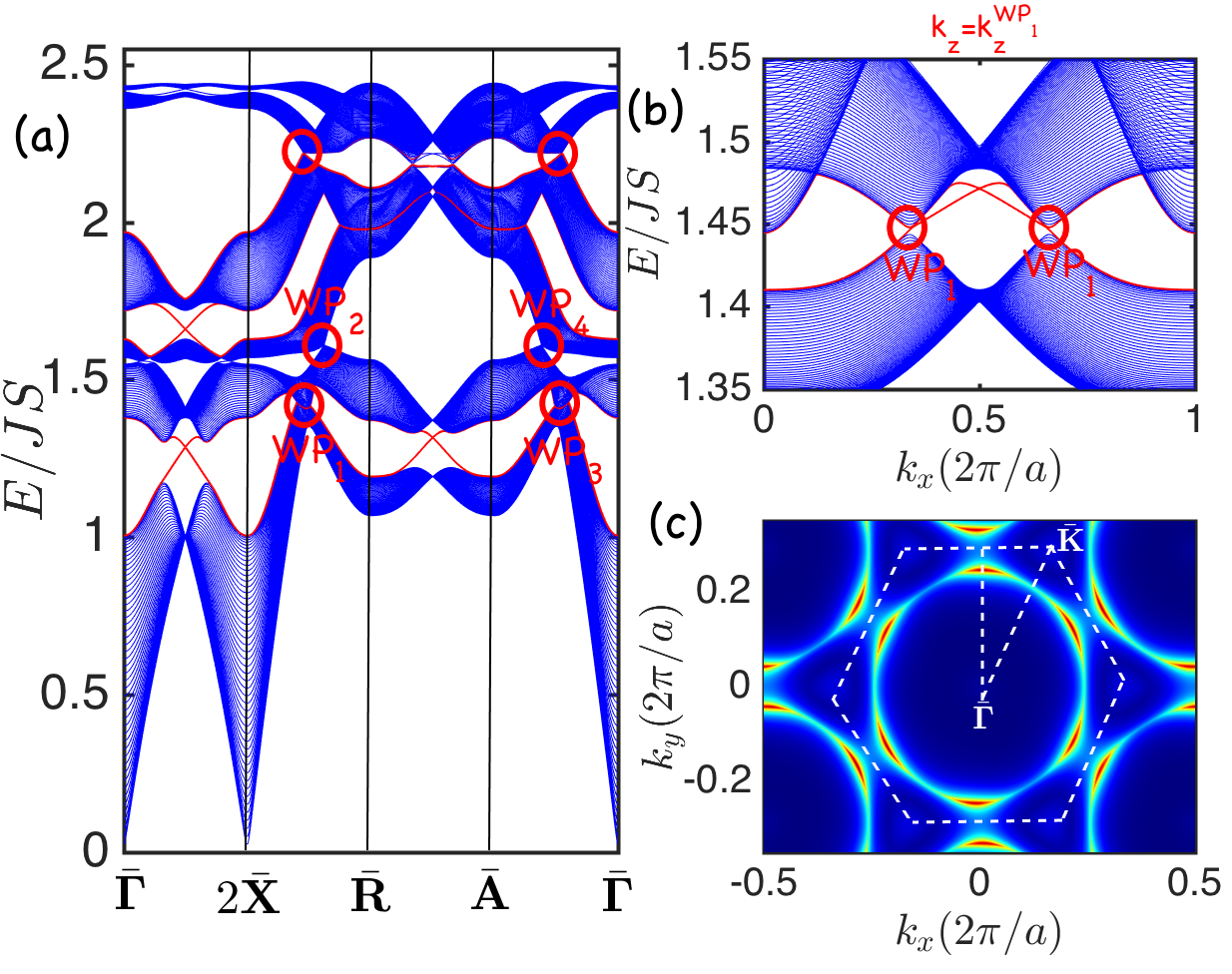}
\caption{Color online.  (a) The projected (010)-surface dispersion along the surface BZ. The red curves are the projected WPs and the red lines indicate the surface states.    (b) The projected (010)-surface dispersion along $k_x$ direction with constant $k_z=k_z^{\text{WP}_1}=0.2626\times 2\pi/c$ at WP$_1$. (c) Magnonic arcs with energy set at WP$_1$  for the (001)-surface state.   The parameters are the same as  in Fig.~\ref{WM}(a).}
\label{WSS}
\end{figure}

\textbf{Magnonic surface state arcs.}-- Besides the bulk magnonic TPs and WPs, another observable quantities are the  magnonic surface states and arcs. They can be found by projecting the bulk magnonic TPs and WPs onto the surface BZs. Let us first consider the projection of the magnonic TPs onto the surface BZ. We  consider the (010) and (001) surfaces. In the former surface, the bulk magnonic TPs along ${\bf \Gamma}$-${\bf A}$ line project onto the $\bar{\bf \Gamma}$-$\bar{\bf A}$ line, and those along ${\bf K}$-${\bf H}$ line project onto the $\bar{\bf X}$-$\bar{\bf R}$  line.   In Fig.~\ref{TSS}(a) we have shown the projected (010)-surface dispersion along the surface BZ,  and Fig.~\ref{TSS}(b) shows the surface dispersion along $k_x$ direction with constant $k_z$ at TP$_1$. We see that the magnonic TPs are connected by a surface state. In Fig.~\ref{TSS}(c) we show the corresponding surface constant energy contours at the energy TP$_1$.  We clearly see that the two bulk magnonic TPs  at TP$_1$ projected onto  $\bar{\bf X}$-$\bar{\bf R}$  line  are connected by a pair of magnonic  arcs on the  (010) surface BZ. In the latter (001)-surface, however, the bulk magnonic TPs along ${\bf \Gamma}$-${\bf A}$ line project onto the $\bar{\bf \Gamma}$ point, and those along ${\bf K}$-${\bf H}$ line project onto the $\bar{\bf K}$ points. Therefore,  on the (001) surface there are no arcs connecting the two projected TPs (not shown).

 Upon breaking of symmetry, the magnonic TPs split into magnonic WPs with only two non-degenerate band crossing points.   In Fig.~\ref{WSS}(a) we have shown the projected (010)-surface dispersion along the surface BZ,  and Fig.~\ref{WSS}(b) shows the surface dispersion along $k_x$ direction with constant $k_z$ at WP$_1$, which shows that two magnonic WP$_1$ are connected by a surface state.  In Fig.~\ref{WSS}(c) we show the corresponding (001)-surface constant energy contours for energy set at WP$_1$.  Because  the two bulk WPs with opposite chirality are projected onto the same point on the (001) surface,  it is clear  that there is no magnonic arc from the projected WPs. However, magnonic arcs exist inside the (001)-surface BZ which connect   two bulk WPs with opposite chirality projected onto the (010)-surface BZ. It is important to note that there are no magnonic surface states or arcs in the strictly 2D system \cite{2d}, therefore the current results are obviously different and require quasi-2D or 3D non-collinear spin structures.

\textbf{Conclusion.}-- In summary, we have introduced the notion of magnetic triply-degenerate nodal points (TPs) in insulating quantum antiferromagnets with non-collinear spin structures. The TPs are formed by the crossing of one doubly-degenerate band and one single non-degenerate band along the high symmetry lines in momentum space, and are protected by magnetic crystal symmetry.  The realization of TPs in insulating quantum antiferromagnets represent quasiparticle excitations with three-component bosons  beyond the band crossing points in magnonic Dirac, Weyl,  and nodal-line cases. We also showed the existence of magnonic arc surface states connecting surface projection of TPs, and the transition of the TPs to Weyl points upon breaking of symmetries.  We note that in metallic antiferromagnets, doubly-degenerate bands can be restored by the combined magnetic crystal and time-reversal symmetry  although time-reversal symmetry itself is broken by the magnetic order. This suggests that TPs formed by the crossing of one doubly-degenerate band and one single non-degenerate band is possible in metallic antiferromagnets. Therefore, we believe that the current study should inspire the search for TPs in metallic quantum antiferromagnets with (non)collinear spin structures.  

\acknowledgments
Research at Perimeter Institute is supported by the Government of Canada through Industry Canada and by the Province of Ontario through the Ministry of Research
and Innovation.


\begin{thebibliography}{0}
\bibitem{foot}
 This Hamiltonian can be obtained by expanding near the magnonic TPs and projecting onto the three magnon bands. The calculation is quite demanding, lengthy, and tedious.
\end{thebibliography}
\end{document}


\date{\today}
\title{ Magnonic triply-degenerate nodal points\\
Supplemental Material}
\author{S. A. Owerre}
\affiliation{Perimeter Institute for Theoretical Physics, 31 Caroline St. N., Waterloo, Ontario N2L 2Y5, Canada.}

\maketitle

\section{  Spin transformation}
Generally, we consider  noncoplanar  chiral spin configurations, which can be induced by the in-plane DMI or an external  magnetic field. In this case there is a spin-canting behaviour out-of-plane leading to a nonzero scalar spin chirality.  In this paper, we consider noncoplanar spin structure induced by an applied external magnetic field along the stacking $z$ direction, $\mathcal H_Z=- H\sum_{i,\ell} S_{i,\ell}^z$ in units of $g\mu_B$. Next,  we perform a rotation about the $z$-axis  by spin orientated angles $\theta_i$ and then about $y$-axis by the magnetic-field-induced spin-canting angle $\eta$. The total rotation matrix is given by
\begin{align}
\mathcal{R}_z(\theta_{i,\ell})\cdot\mathcal{R}_y(\eta)
=\begin{pmatrix}
\cos\theta_{i,\ell}\cos\eta & -\sin\theta_{i,\ell} & \cos\theta_i\sin\eta\\
\sin\theta_{i,\ell}\cos\eta & \cos\theta_{i,\ell} &\sin\theta_{i,\ell}\sin\eta\\
-\sin\eta & 0 &\cos\eta
\end{pmatrix}.
\end{align}
 Consequently, the spins transform as $ \bold{S}_i=\mathcal{R}_z(\theta_{i,\ell})\cdot\mathcal{R}_y(\eta)\cdot\bold S_i^\prime,$
where prime denotes spins in the rotated frame. The classical ground state energy  is given by \begin{align}
E_{\text{cl}}&= 6NS^2\Big[-\frac{J_T}{2}\lb 1 - {3}\cos^2\eta\rb-\frac{J_D}{2}\lb 1 - 2\cos^2\eta\rb\\&\nonumber-\frac{\sqrt{3}}{2}D_z\sin^2\eta -\frac{J_\perp}{2}(1-2\cos^2\eta)-H\cos\eta\Big],
\end{align}
where $N$ is the number of sites per unit cell, and the magnetic field is rescaled in unit of $S$. Minimizing this energy yields the canting angle $\cos\eta = H/H_s$, where $H_s=3J_T+2J_D+\sqrt{3}D_z+ 2J_\perp$ is the saturation field. The spin interactions  contributing to the free magnon model (in terms of the  Holstein-Primakoff  bosons) are  given by

  \begin{align}
  \mathcal H_J&= \sum_{ \la ij\ra,\ell}J_{ij}\big[\cos\theta_{ij,\ell} \bold{ S}_{i,\ell}^\prime\cdot \bold{ S}_{j,\ell}^\prime+ \sin\theta_{ij,\ell}\cos\eta \hat{\bold z}\cdot\lb \bold{ S}_{i,\ell}^\prime\times\bold{ S}_{j,\ell}^\prime\rb \\&\nonumber+2\sin^2\lb\frac{\theta_{ij,\ell}}{2}\rb\lb\sin^2\eta  S_{i,\ell}^{\prime x}S_{j,\ell}^{\prime x} +\cos^2\eta S_{i,\ell}^{\prime z} S_{j,\ell}^{\prime z}\rb\big],
  \end{align}
  
 \begin{align}
\mathcal   H_{D_z}&= -D_z\sum_{\la ij\ra,\ell}\Big[\cos\theta_{ij,\ell}\cos\eta {\bf \hat z}\cdot \lb \bold S_{i,\ell}^\prime\times\bold S_{j,\ell}^\prime\rb  \\&\nonumber - \sin\theta_{ij,\ell}\lb \cos^2\eta S_{i,\ell}^{\prime x}S_{j,\ell}^{\prime x} + S_{i,\ell}^{\prime y}S_{j,\ell}^{ \prime y}+\sin^2\eta S_{i,\ell}^{\prime z}S_{j,\ell}^{\prime z}\rb\Big], 
     \end{align}
 
    \begin{align}
  \mathcal H_{J_\perp}&= J_\perp\sum_{ i,\la \ell\ell^\prime\ra}\big[\cos\theta_{\ell\ell^\prime} \bold{ S}_{i,\ell}^\prime\cdot \bold{ S}_{i,\ell^\prime}^\prime  \\&\nonumber  +2\sin^2\lb\frac{\theta_{\ell\ell^\prime}}{2}\rb\lb\sin^2\eta  S_{i,\ell}^{\prime x}S_{i,\ell^\prime}^{\prime x} +\cos^2\eta S_{i,\ell}^{\prime z} S_{i,\ell^\prime}^{\prime z}\rb\big],
  \label{rotp}
  \end{align}
  \begin{align}
 \mathcal H_Z= -H\cos\eta\sum_{i,\ell} S_{i,\ell}^{\prime z},
  \end{align}
where $\theta_{\alpha\beta}=\theta_{\alpha}-\theta_{\beta}$. For antiferromagnetic interlayer coupling $J_\perp>0$ we have $\theta_{\ell\ell^\prime}=\pi$, whereas  $\theta_{\ell\ell^\prime}=0$ for ferromagnetic interlayer $J_\perp<0$. In this latter case  only the first term in  Eq.~\eqref{rotp} is nonzero. In addition, $\theta_{ij}=\pi$ on the dimer bonds $J_D$, whereas $\theta_{ij}=2\pi/3$ on the triangular  bonds $J_T$.  The noncoplanar  (umbrella) spin configuration has a nonzero scalar spin chirality given by $\chi_{ijk;l}=  {\bf S}_{i,\ell}^\prime\cdot({\bf S}_{j,\ell}^\prime\times{\bf S}_{k,\ell}^\prime)$, and it is induced only within the decorated honeycomb lattice  planes. 
 \begin{figure}
\centering
\includegraphics[width=1\linewidth]{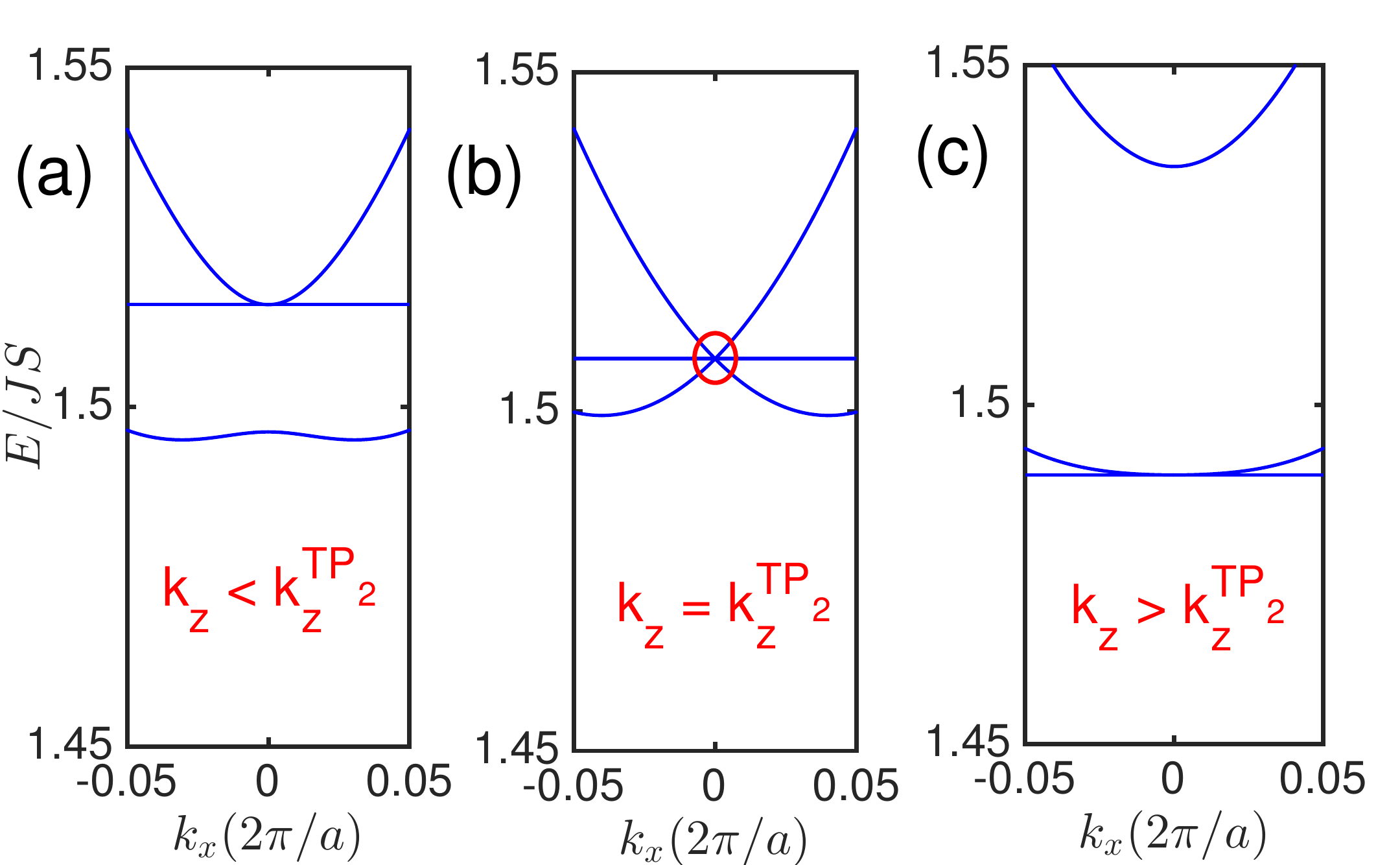}
\caption{Color online. Magnonic TP phase below (a), at (b), and above (c) TP$_2$  for fixed  $k_z^{\text{TP}_2}=1.6520\times 2\pi/c$. The parameters are the same as  in Fig.~\ref{TDNLM}(a).}
\label{TP}
\end{figure} 
\section{Holstein-Primakoff  bosons}

Next, we introduce the  Holstein-Primakoff  bosons: $
S_{i,\ell}^{z}= S-a_{i,\ell}^\dagger a_{i,\ell},~  S_{i,\ell}^{+} \approx  \sqrt{2S}a_{i,\ell}=\lb S_{i,\ell}^{-}\rb^\dg$, where $ S_{i,\ell}^{\pm}=S_{i,\ell}^{x}\pm i S_{i,\ell}^{y}$ and $a_{i,\ell}^\dagger(a_{i,\ell})$ are the bosonic creation (annihilation) operators. In the following we consider the case $J_\perp>0$. The case $J_\perp<0$ can be derived in a similar way. The magnon hopping Hamiltonians are given by
\begin{align}
\mathcal H_{J_T-D_z}&= S\sum_{\la ij\ra,\ell}\big[ t^z(a_{i,\ell}^\dg a_{i,\ell} +a_{j,\ell}^\dg a_{j,\ell}) \\&\nonumber + t^r(e^{-i\phi_{ij,\ell}}a_{i,\ell}^\dg a_{j,\ell} + h.c.)+ t^o(a_{i,\ell}^\dg a_{j,\ell}^\dg + h.c.)\big],
\end{align}
\begin{align}
\mathcal H_{J_D}&= S\sum_{\la ij\ra,\ell}\big[ t^{z}_D(a_{i,\ell}^\dg a_{i,\ell} +a_{j,\ell}^\dg a_{j,\ell}) \\&\nonumber + t^{r}_D(a_{i,\ell}^\dg a_{j,\ell} + h.c.)+ t^{o}_D(a_{i,\ell}^\dg a_{j,\ell}^\dg + h.c.)\big],
\end{align}
 \begin{align}
\mathcal H_{J_\perp}&= S\sum_{i,\ell} t_\perp^za_{i,\ell}^\dg a_{i,\ell}  \\&\nonumber + S\sum_{i,\la \ell\ell^\prime\ra}\big[ t_\perp^r(a_{i,\ell}^\dg a_{i,\ell^\prime} + h.c.) + t_\perp^o(a_{i,\ell}^\dg a_{i,\ell^\prime}^\dg + h.c.)\big],
\end{align}
\begin{align}
\mathcal H_{Z}&=H\cos\eta\sum_{i,\ell}a_{i,\ell}^\dg a_{i,\ell}.
\end{align}
   The  solid angle subtended by three non-coplanar spins  is given by $\phi_{ij}=\pm\phi$, where $\phi=\tan^{-1}[t_{2}^r/t_{1}^r]$. The parameters of the tight binding model are given by 

\begin{align}
&t^z= -J_T\lb -\frac{1}{2}+\frac{3}{2}\cos^2\eta\rb+\frac{ \sqrt{3}D_z}{2}\sin^2\eta,\\
&t^r=\sqrt{(t_1^r)^2+(t_2^r)^2},\\
&t_1^r= J_T\lb-\frac{1}{2} +\frac{3\sin^2\eta}{4}\rb-\frac{\sqrt{3}D_z}{2}\lb 1-\frac{\sin^2\eta}{2}\rb,\\
&t_2^r= \cos\eta\lb -\frac{\sqrt{3}J_T}{2}+\frac{D_z}{2}\rb,\\
&t^{o}=\frac{\sin^2\eta}{2}\lb \frac{3J_T}{2}+\frac{\sqrt{3}D_z}{2}\rb,\\
&t^{z}_D= \frac{J_D}{2}\lb -1+2\cos^2\eta\rb,\\
&t^r_D= J_D\lb-1 +\sin^2\eta\rb,\\
& t^{o}_D=J_D\sin^2\eta,\\
& t_\perp^z= -\frac{J_{\perp}}{2}\cos 2\eta,~
t_\perp^r=-  J_{\perp}\cos 2\eta,~
t_\perp^o=J_{\perp} \sin^2\eta.
\end{align}

\section{Magnon Hamiltonian}

Next, we Fourier transform into momentum space. The resulting  Hamiltonian 
using the  basis vector $\Psi^\dg_\bo=(\psi^\dg_\bo,\psi^\dg_{-\bo})$, where $\psi^\dg_\bo=(a_{\bo 1}^{\dg},\thinspace a_{\bo 2}^{\dg},\thinspace a_{\bo 3}^{\dg}, \thinspace a_{\bo 4}^{\dg},\thinspace a_{\bo 5}^{\dg},\thinspace a_{\bo 6}^{\dg} )$, is given by 
\begin{align}
\boldsymbol{\mathcal{H}}(k_z,k_\parallel)= 
\begin{pmatrix}
\boldsymbol{\mathcal{A}}(k_z,k_\parallel, \phi) & \boldsymbol{\mathcal{B}}(k_z,k_\parallel)\\
\boldsymbol{\mathcal{B}}^*(-k_z,-k_\parallel)&\boldsymbol{\mathcal{A}}^*(-k_z,-k_\parallel,\phi)
\end{pmatrix},
\label{ham}
\end{align}
where 
\begin{align}
\boldsymbol{\mathcal{A}}(k_z,k_\parallel)= &
\begin{pmatrix}
\bold{A}_1(k_z,\phi) & \bold{A}_2(k_\parallel)\\
\bold{A}_2(-k_\parallel)&\bold{A}_1(k_z,\phi)
\end{pmatrix},\\
\boldsymbol{\mathcal{B}}(k_z,k_\parallel)= &
\begin{pmatrix}
\bold{B}_1(k_z) & \bold{B}_2(k_\parallel)\\
\bold{B}_2(-k_\parallel)& \bold{B}_1(k_z)
\end{pmatrix},
\end{align}
 
\begin{align}
\bold{A}_1(k_z,\phi)=& 
\begin{pmatrix}
t_0(k_z)&t^r e^{-i\phi}&t^r e^{i\phi}\\
t^r e^{i\phi}&t_0(k_z)&t^r e^{-i\phi}\\
t^r e^{-i\phi}&t^r e^{i\phi}&t_0(k_z)
\end{pmatrix},\\
\bold{A}_2(k_\parallel)=&t^{r}_D
\begin{pmatrix}
 e^{ik_{2\parallel}}&0&0\\
0& e^{ik_{1\parallel}}&0\\
0&0&1
\end{pmatrix},
\end{align}
\begin{align}
\bold{B}_1(k_z)= &
\begin{pmatrix}
t_\perp^o\cos k_z c&t^o &t^o \\
t^o &t_\perp^o\cos k_z c&t^o \\
t^o &t^o &t_\perp^o\cos k_z c
\end{pmatrix},\\
\bold{B}_2(k_\parallel)=&t^o_D
\begin{pmatrix}
e^{ik_{2\parallel}}&0&0\\
0& e^{ik_{1\parallel}}&0\\
0&0&1\\
\end{pmatrix},
\end{align}
where $t_0(k_z)=H\cos\eta+2(t^z + t^{z}_D+t_\perp^{z})=J_T+J_D +\sqrt{3}D_z+J_\perp+t_\perp^r\cos k_z c$ and $k_{i\parallel}=k_\parallel\cdot \bold{e}_{i}$, with  $k_\parallel=(k_x,k_y)$ and  $\bo=(k_\parallel,k_z)$. The lattice basis vectors are chosen as $\bold e_1=2a\hat{x}$ and  $\bold e_2=a(\hat{x} + \sqrt{3}\hat{ y})$.  The magnon band structure below, at, and above TP$_2$ are shown in Fig.~\ref{TP}.